\documentclass[aps,pra, twocolumn, superscriptaddress,amsmath,amssymb,floatfix]{revtex4}
\usepackage{graphicx}
\usepackage{times}
\usepackage{caption}
\usepackage{graphicx,amsmath,amssymb,amstext,amsfonts,color}

\usepackage{amsmath}
\usepackage{amssymb}

\begin{document}

\title{Quasi-Species in High Dimensional Spaces}
\author{Vaibhav Madhok}
\email{vmadhok@gmail.com}
\affiliation{Department of Zoology and Department of Mathematics, University of British Columbia, Vancouver}

%\author{Michael Doebeli}

%\author{C}

\def\T{\Theta}
\def\D{\Delta}
\def\d{\delta}
\def\r{\rho}
\def\p{\pi}
\def\a{\alpha}
\def\g{\gamma}
\def\ra{\rightarrow}
\def\s{\sigma}
\def\b{\beta}
\def\e{\epsilon}
\def\G{\Gamma}
\def\om{\omega}
\def\l{\lambda}
\def\f{\phi}
\def\w{\psi}
\def\m{\mu}
\def\t{\tau}
\def\c{\chi}
\begin{abstract}
We show that, under certain assumptions, the fitness of almost all quasi-species becomes independent of mutational probabilities and the initial frequency distributions of the sequences in high dimensional sequence spaces. This result is the consequence of the concentration of measure on a high dimensional hypersphere and its extension to Lipschitz functions knows as the Levy's Lemma. Therefore, evolutionary dynamics almost always yields the same value for fitness of the quasi-species, independent of the mutational process and initial conditions, and is quite robust to mutational changes and fluctuations in initial conditions. Our results naturally extend to any Lipschitz function whose input parameters are the frequencies of individual constituents of the quasi-species. This suggests that the functional capabilities of high dimensional quasi-species are robust to fluctuations in the mutational probabilities and initial conditions.

\end{abstract}

\maketitle

\vskip 0.5 cm
%begin{twocolumn}
\section{Introduction} 
Living systems and life processes show a remarkable order despite the role of chance and mutational
processes underlying its origin.
Why are living systems so well adapted to their environment?  The slogan, ``survival of the fittest" seems to be the accepted answer. However, without a proper definition of ``fitness", the question of adaptation remains unanswered. Indeed, without a proper quantification of fitness, the above argument reduces to ``survival of the survivors",  which by its circular nature is an uncomfortable position to study evolutionary biology.

The subject of ``fitness" as a quantity that is optimized by evolution is a contentious issue in evolutionary theory. One the one hand, the performance of biological systems, characterized as physical processes, is near optimal and close to the limits set by the laws of physics. In the domain of biochemical processes, the  enzymes for example, serve as nearly optimal catalysts.
On the other hand,  evolution viewed as a complex dynamical process with mutations, its essential fuel, being stochastic in nature, probably does not optimize anything. 
The quasi-species model \cite{eigen77, eigen71, fontana87, swetina82, McCaskill84, nowak89, eigen89} is an attempt to address these questions.
%Another important question is whether it is nevertheless possible to make general statements about evolving systems despite the debate around ``fitness". 
%Our study is an attempt to answer these questions using results from high dimensional geometry.

One of the motivations behind the concept of quasi-species, introduced by Eigen and Schuster, was to be able to make precise statements about the notion of the survival of the fittest \cite{eigen77, eigen71}.  Quasi-Species is an ensemble with a well defined distribution of mutants that is a result of the evolutionary process involving selection and mutation. Selection acts on the quasi-species as a whole and the most optimal ensemble survives.
Quasi-Species sheds light on the role of chance in the process of adaptation by taking into account the role of errors in the process of replication which results in the generation of an ensemble of closely related species instead of a single fittest constituent. The equilibrium distribution resulting from the selection mutation process is depends not only on the replication rates of individual constituents but also on the erroneous replication of the entire population.
Therefore, natural selection as an optimization is not directed toward the single fittest variant, but towards the ensemble which evolve to maximize its average replication rate. In general, the average replication rate, 
also known as the mean fitness will depend on the relative frequencies of the variants which in turn depends on the underlying mutational probabilities. Therefore, while the quasi-species formulation shows the role of chance in the process of adaptation, the near optimal adaptation observed is often attributed to the ``fitness" of the whole quasi-species.
After all, as mentioned above, living systems including biochemical processes like enzyme functions show efficient adaptation regardless of what role chance might have had. Moreover, the mutational probabilities that cause cross-coupling between the individual variants have their origins in quantum mechanics and in general, should not be assumed to be fixed in the entire course of evolution. Our work shows the robustness of the fitness function even when these assumptions are relaxed. Therefore, for almost all mutational rates and initial conditions, the resulting quasi-species at equilibrium are equally fit and more importantly show quantitatively similar functional capabilities.
Ours is a first example of quasispecies in high dimensional phenotypic spaces.....   
   
\section{Quasi-Species Equation}
Quasi-Species is an ensemble with a well defined distribution of mutants that is a result of the evolutionary process involving selection and mutation. Selection acts on the quasi-species as a whole and the most optimal ensemble survives  
  
  Quasi-Species is an ensemble of related sequences

 \begin{align}
\label{QS1}
 \frac{d X}{dt} = W X - g(X). X
\end{align}
 The vector $X$ consists of the population densities of the individual sequences, 
  \begin{align}
\label{QS2}
 X = (x_1, x_2,..., x_n)
\end{align}
  
  The matrix $W$ consists of individual replication rates, $a_{i}, i = 1, 2, ..., n$, along with the mutation rates for transition between individual sequences, $i$ and $j$, given by $Q_{ij}$.
  
\begin{align}
 W = \begin{pmatrix}a_1 Q_{11}&a_2Q_{12}&...&a_2Q_{1n} \\ a_1 Q_{21}&a_2Q_{22}&...&a_2Q_{2n}\\...&...&...&...\\a_1 Q_{n1}&a_2Q_{n2}&...&a_2Q_{nn} \end{pmatrix}_.
\end{align}  
  
  The total size of the population is a constant if we have 
  
  \begin{align}
\label{QS3}
g(X) = \sum_{i=1}^{n} a_{i}x_{i} / \sum_{i}^{n} x_{i}
\end{align}.
  
The equilibrium of Eq. \ref{QS1} is given by solving the eigenvalue problem
\begin{align}
\label{QS3}
WX = \lambda X
\end{align}
  The fact that the above system will have a unique largest positive eigenvalue is guaranteed by the Frobenius Perron \cite{perron07, frobenius12} theorem.
  
The largest eigenvalue gives the average replication rate of the quasi-species, $\lambda_{max} = \sum_{i=1}^{n} a_{i}x_{i}$ and the corresponding eigenvector gives the frequency distribution, $X_{eq} = ({x_{1},x_{2},..., x_{n}})$, at equilibrium.

\subsection{Application of Levy's Lemma to Quasi-Species}  
   We now introduce the Levy's Lemma \cite{levy} and apply it to the solution of the quasi-species equation.

\textbf{Levy's Lemma:} % (Levy. See [19], Appendix IV, and [15]). 
Let $f : S^k  \rightarrow R$
be a function with Lipschitz constant $\eta$ (with respect to the Euclidean norm) and $X \in k$
and a point $X \in S^{k}$ be chosen uniformly at random. Then
\begin{align}
\label{levy}
 Pr\{|f(X) - \bar{f}| >  \alpha\}\le exp (-C(k+1)\alpha^{2}/\eta^2 )
\end{align}
for some constant $C >0$ and $\bar{f}$ is the mean value of the function over the sphere.
 
\textbf{Mapping the Quasi-Species on a hypersphere} 
 
 One can map the state describing the quasi-species on the surface of an $n$ dimensional hypersphere.
 The coordinates, $Y =(y_1, y_2, ..., y_n$), of the hypersphere are given by ($ {\sqrt{x_1}, \sqrt{x_2}, ..., \sqrt{x_n}}$).
The function, $f$, relevant to us, is the mean fitness or the average replication rate $f=\sum_{i=1}^{n} a_{i}x_{i} = \sum_{i=1}^{n} a_{i}y_{i}^2$.  A function $h$ is Lipschitz, if, 
\begin{align}
\label{lipchitz1}
 |h(X) - h(Y)| \leq C|X-Y|,
\end{align}
for all $X$ and $Y$ and C is a constant independent of $X$ and $Y$. The fitness function, $f$, is Lipschitz as $\nabla{f}$ is bounded by  $2\max_{1 \leq j \leq n}\{a_j\} $.

 %$\bar{w}(y_1^2,...,y_n^2)$
%We can compute any \textit{reasonable} function on the surface of the hypersphere.
 Therefore, Levy's Lemma shows us that for any point picked at random on a high dimensional hypersphere, 
 the value of the fitness function will be concentrated around the mean, $\bar{f}$, with high probability.
Therefore, a random point on the hypersphere represents the state of the quasi-specie at a given time.
Levy's lemma shows that almost all quasi-species in higher dimensions have the same fitness.
In the above, we have assumed, $Y$, to be uniformly random over the sphere.

\subsection{On the robustness of functional capabilities of quasi-species} 
 Any function, G, whose input parameters are the frequencies of the individual sequences, $X$, can be computed for the distribution given by the eigenvectors is concentrated closely about its average over the entire hypersphere. Therefore, if the functional behavior of quasi-species is given by such a function, its value is independent of the mutational matrix and the initial conditions. This also suggests robustness of functional behavior of quasi-species to perturbations in mutation rates and initial conditions. This is a significant result as one might expect that the workings of certain life processes, as described by quasi-species, require a certain degree of accuracy and robustness which we have shown is possible in high dimensional spaces \cite{schro}.

 \section{Conclusion}
 We have shown that the fitness of quasi-species is kinematical in nature, i.e. dependent on the system dimensions and individual selection rates and independent of the mutation dynamics and initial conditions.
  For almost all initial quasi-species distributions and mutation error probabilities, evolution leads to almost the same value for fitness fitness as defined by the largest eigenvalue of the mutation-selection matrix.
We have also shown how the functional capabilities of quasi-species is robust to mutational changes and fluctuations in the initial conditions. Our work is a consequence of application of ideas from high dimensional geometry to demonstrate the robustness of certain life processes and should be of use to explore the questions related to the origin of life.

\end{document}